\documentclass[12pt,fleqn]{article}
\usepackage{graphicx,cite}
\usepackage[T1]{fontenc}
\usepackage{amsfonts}
\usepackage{multirow}
\usepackage{amsthm}
\usepackage{amsmath}
\usepackage[utf8]{inputenc}
\newcommand{\nc}{\newcommand}

\nc{\newsection}[1]{\section{#1}\setcounter{equation}{0}}
\nc{\newappendix}[1]{\section*{#1}\setcounter{equation}{0}}
\def\lb{\linebreak}
\def\nnb{\nonumber}
\nc{\scm}{\scriptscriptstyle\mathrm}
\nc{\f}{\frac}
\nc{\be }{\begin{equation}}   \nc{\ee }{\end{equation}}
\nc{\bea}{\begin{eqnarray}}   \nc{\eea}{\end{eqnarray}}
\nc{\baa}{\begin{array}}      \nc{\eaa}{\end{array}}
\nc{\bit}{\begin{itemize}}    \nc{\eit}{\end{itemize}}
\nc{\ben}{\begin{enumerate}}  \nc{\een}{\end{enumerate}}
\nc{\bce}{\begin{center}}     \nc{\ece}{\end{center}}
\nc{\bfl}{\begin{flushright}} \nc{\efl}{\end{flushright}}
\nc{\btb}{\begin{tabular}}    \nc{\etb}{\end{tabular}}
\nc{\eps}{\varepsilon}
\nc{\vp}{\varphi}
\nc{\tvp}{\widetilde{\varphi}}
\nc{\D}{\mbox{$\not\!\!D$}}
\nc{\Db}{\mbox{${\raisebox{2mm}{\boldmath ${}^\leftarrow$}\hspace{-4mm} D}$}}
\nc{\Dfb}{\mbox{$\raisebox{2mm}{\boldmath ${}^\leftrightarrow$}\hspace{-4mm} D$}}
\nc{\vpj }{\mbox{${\vp^\dag i\,\raisebox{2mm}{\boldmath ${}^\leftrightarrow$}\hspace{-4mm} D_\mu\,\vp}$}}
\nc{\vpjt}{\mbox{${\vp^\dag i\,\raisebox{2mm}{\boldmath ${}^\leftrightarrow$}\hspace{-4mm} D_\mu^{\,I}\,\vp}$}}
\def\ocal{{\cal O}}
\def\lcal{{\cal L}}
\def\p{\partial}
\def\wt{\widetilde}
\begin{document}
\begin{titlepage}
\bfl \btb{l}
IFT-9/2010\\
TTP10-35\\[25mm]
\etb \efl
\bce
\setlength {\baselineskip}{0.3in} 
{\bf\Large Dimension-Six Terms in the Standard Model Lagrangian$^\star$}\\[2cm]
\setlength {\baselineskip}{0.2in}
{\large  B.~Grzadkowski$^1$, M.~Iskrzy\'nski$^1$, M.~Misiak$^{1,2}$~ and~ J.~Rosiek$^1$}\\[5mm]
$^1$~{\it Institute of Theoretical Physics, University of Warsaw,\\
         Ho\.za 69, PL-00-681 Warsaw, Poland.}\\[5mm] 
$^2$~{\it Institut f\"ur Theoretische Teilchenphysik, 
          Karlsruhe Institute of Technology (KIT),\\
          D-76128 Karlsruhe, Germany.}\\[2cm]
{\bf Abstract}\\[5mm]
\ece
\setlength{\baselineskip}{0.2in} 

When the Standard Model is considered as an effective low-energy theory,
higher dimensional interaction terms appear in the Lagrangian. Dimension-six
terms have been enumerated in the classical article by Buchm\"uller and
Wyler~\cite{Buchmuller:1985jz}. Although redundance of some of those operators
has been already noted in the literature, no updated complete list has been
published to date. Here we perform their classification once again from the
outset. Assuming baryon number conservation, we find $15+19+25=59$ independent
operators (barring flavour structure and Hermitian conjugations), as compared
to $16+35+29=80$ in Ref.~\cite{Buchmuller:1985jz}. The three summed numbers
refer to operators containing 0, 2 and 4 fermion fields.  If the assumption of
baryon number conservation is relaxed, 4 new operators arise in the four-fermion
sector.

\ \\[2cm]
\setlength {\baselineskip}{0.2in}
\noindent \underline{\hspace{2in}}\\
\noindent
$^\star$ {\footnotesize This paper is based on the M.Sc. thesis of the second author.}

\end{titlepage}

\newsection{Introduction \label{sec:intro}}

The Standard Model (SM) of strong and electroweak interactions has been
successfully tested to a great precision~\cite{Nakamura:2010XXX}. Nevertheless,
it is commonly accepted that it constitutes merely an effective theory which
is applicable up to energies not exceeding a certain scale $\Lambda$. A field
theory valid above that scale should satisfy the following requirements: 
\bit
\item[({\it i})] its gauge group should contain~ $SU(3)_C\times SU(2)_L\times U(1)_Y$~
of the SM,
\item[({\it ii})] all the SM degrees of freedom should be incorporated
either as fundamental or composite fields,
\item[({\it iii})] at low-energies, it should reduce to the SM, provided no
  undiscovered but weakly coupled {\em light} particles exist, like
  axions or sterile neutrinos.
\eit

In most of beyond-SM theories that have been considered to date, reduction to
the SM at low energies proceeds via decoupling of heavy particles with masses
of order $\Lambda$ or larger. Such a decoupling at the perturbative level is
described by the Appelquist-Carazzone theorem~\cite{Appelquist:1974tg}. This
inevitably leads to appearance of higher-dimensional
operators in the SM Lagrangian that are suppressed by
powers of $\Lambda$
\be \label{Leff}
\lcal_{\mathrm SM} = \lcal_{\mathrm SM}^{(4)} + \f{1}{\Lambda  } \sum_{k} C_k^{(5)} Q_k^{(5)}
+ \f{1}{\Lambda^2} \sum_{k} C_k^{(6)} Q_k^{(6)} + \ocal\left(\f{1}{\Lambda^3}\right), 
\ee
where $\lcal_{\mathrm SM}^{(4)}$ is the usual ``renormalizable'' part of the SM
Lagrangian. It contains dimension-two and -four operators only.\footnote{
  Canonical dimensions of operators are determined from the field contents alone,
  excluding possible dimensionful coupling constants.  The only dimension-two
  operator in~ $\lcal_{\mathrm SM}^{(4)}$~ is~ $\vp^\dag \vp$~ in the Higgs mass term.}
In the remaining terms, $Q_k^{(n)}$ denote dimension-$n$ operators, and
$C_k^{(n)}$ stand for the corresponding dimensionless coupling constants
(Wilson coefficients). Once the underlying high-energy theory is specified,
all the coefficients $C_k^{(n)}$ can be determined by integrating out the
heavy fields.

Our goal in this paper is to find a complete set of independent operators of
dimension 5 and 6 that are built out of the SM fields and are consistent with
the SM gauge symmetries. We do not rely on the original analysis of such
operators by Buchm\"uller and Wyler~\cite{Buchmuller:1985jz} but rather
perform the full classification once again from the outset. One of the reasons
for repeating the analysis is the fact that many linear combinations of
operators listed in Ref.~\cite{Buchmuller:1985jz} vanish by the Equations Of
Motion (EOMs). Such operators are redundant, i.e. they give no contribution to
on-shell matrix elements, both in perturbation theory (to all orders) and
beyond~\cite{Politzer:1980me,KlubergStern:1975hc,GrosseKnetter:1993td,Arzt:1993gz,Simma:1993ky,Wudka:1994ny}.
Although the presence of several EOM-vanishing combinations in
Ref.~\cite{Buchmuller:1985jz} has been already pointed out in the
literature~\cite{Grzadkowski:2003tf,Fox:2007in,AguilarSaavedra:2008zc,AguilarSaavedra:2009mx},
no updated complete list has been published to date. Our final operator basis
differs from Ref.~\cite{Buchmuller:1985jz} also in the four-fermion sector
where the EOMs play no role.

The article is organized as follows. Our notation and conventions are
specified in Sec.~\ref{sec:not}. The complete operator list is presented in
Sec.~\ref{sec:results}. Comparison with Ref.~\cite{Buchmuller:1985jz} is
outlined in Sec.~\ref{sec:compar}.  Details of establishing operator
bases in the zero-, two- and four-fermion sectors are described in
Secs.~\ref{sec:bosons}, \ref{sec:twoferm} and \ref{sec:fourferm},
respectively. We conclude in Sec.~\ref{sec:concl}.

\newsection{ Notation and conventions\label{sec:not}}

The SM matter content is summarized in Tab.~\ref{tab:matter} with isospin,
colour, and generation indices denoted by~ $j=1,2$,~ $\alpha=1,2,3$,~ and~
$p=1,2,3$,~ respectively.  Chirality indices ($L$, $R$) of the fermion fields
will be skipped in what follows. Complex conjugate of the Higgs field will
always occur either as $\vp^\dag$ or $\tvp$, where $\tvp^j = \eps_{jk}
(\vp^k)^\star$, and $\eps_{jk}$ is totally antisymmetric with $\eps_{12}=+1$.
\begin{table}[t]
\bce \btb{|c|ccccc|c|}
\hline &&&&&&\\[-3mm]
& \multicolumn{5}{|c|}{fermions} & scalars \\[1mm]\hline&&&&&&\\[-3mm]
field & $l^j_{Lp}$ & $e_{Rp}$ & $q^{\alpha j}_{Lp}$ & $u^\alpha_{Rp}$ & $d^\alpha_{Rp}$ & $\vp^j$ \\[1mm]
\hline&&&&&&\\[-3mm]
hypercharge $Y$ & $-\f12$ & $-1$ & $\f16$ & $\f23$ & $-\f13$ & $\f12$ \\[1mm]
\hline 
\etb \ece
\caption{\sf The SM matter content \label{tab:matter}}
\end{table}

The well-known expression for $\lcal_{\mathrm SM}^{(4)}$ 
before Spontaneous Symmetry Breakdown (SSB) reads
\bea \label{SMlag4}
\lcal_{\mathrm SM}^{(4)} &=& -\f14 G_{\mu\nu}^A G^{A\mu\nu}
                   -\f14 W_{\mu\nu}^I W^{I\mu\nu}
                   -\f14 B_{\mu\nu}   B^{\mu\nu}
   + \left( D_\mu \vp \right)^\dag \left( D^\mu \vp \right) 
   + m^2 \vp^\dag \vp -\f12 \lambda \left( \vp^\dag \vp \right)^2\nnb\\[1mm]
&+& i \left( \bar l \D l + \bar e \D e + \bar q \D q + \bar u \D u + \bar d \D d \right)
 -\left( \bar l\,\Gamma_{\!e} e \vp+ \bar q\,\Gamma_{\!u} u \tvp +
 \bar q\,\Gamma_{\!d} d \vp +\mathrm{h.c.}\right),
\eea
where the Yukawa couplings $\Gamma_{\!e,u,d}$ are matrices in the generation space.
We shall not consider SSB in this paper.
Our sign convention for covariant derivatives is exemplified by
\be \label{D-sign}
\left( D_\mu q \right)^{\alpha j} = \left[ \delta_{\alpha\beta} \delta_{jk} \left( \p_\mu + i g' Y_q B_\mu \right)
+ i g \delta_{\alpha\beta} S^I_{jk} W^I_\mu + i g_s \delta_{jk} T^A_{\alpha\beta} G^A_\mu \right]  q^{\beta k}. 
\ee
Here, $T^{A}=\f12 \lambda^A$ and $S^{I}=\f12 \tau^I$ are the $SU(3)$ and $SU(2)$
generators, while $\lambda^A$ and $\tau^I$ are the Gell-Mann and Pauli
matrices, respectively. All the hypercharges $Y$ have been listed in
Tab.~\ref{tab:matter}.

It is useful to define Hermitian derivative terms that contain $\vp^\dag \Db_\mu\vp
  \equiv (D_\mu\vp)^\dag\vp$ as follows:
\be \label{Dfb}
\vpj \equiv i \vp^\dag \left( D_\mu - \Db_\mu \right) \vp 
\mbox{\hspace{5mm} and \hspace{5mm}} 
\vpjt \equiv i \vp^\dag \left( \tau^I D_\mu - \Db_\mu \tau^I \right) \vp.
\ee

The gauge field strength tensors and their covariant derivatives read
\mathindent0cm
\bea 
\baa{rclcrcl}
G_{\mu\nu}^A &=& \p_\mu G_\nu^A - \p_\nu G_\mu^A - g_s f^{ABC} G_\mu^B G_\nu^C, &\hspace{5mm}&
\left(D_\rho G_{\mu\nu}\right)^A &=& \p_\rho G_{\mu\nu}^A - g_s f^{ABC} G_\rho^B G_{\mu\nu}^C, \\[2mm]
W_{\mu\nu}^I &=& \p_\mu W_\nu^I - \p_\nu W_\mu^I - g \eps^{IJK} W_\mu^J W_\nu^K, &&
\left(D_\rho W_{\mu\nu}\right)^I &=& \p_\rho W_{\mu\nu}^I - g \eps^{IJK} W_\rho^J W_{\mu\nu}^K,\\[2mm] 
B_{\mu\nu}   &=& \p_\mu B_\nu - \p_\nu B_\mu, &&
D_\rho B_{\mu\nu}   &=& \p_\rho B_{\mu\nu}.
\eaa\nnb\\[-7mm] 
\eea
\mathindent1cm
Dual tensors are defined by~ 
$\wt X_{\mu\nu} = \f12 \eps_{\mu\nu\rho\sigma} X^{\rho\sigma}$~
($\eps_{0123}=+1$), where $X$ stands for $G^A$, $W^I$ or $B$. 

The fermion kinetic terms in $\lcal_{\mathrm SM}^{(4)}$ are Hermitian up to total
derivatives, i.e.
$i\bar\psi\D\psi - \mathrm{h.c.}\lb = \p_\mu (\bar\psi \gamma^\mu \psi)$.~
Total derivatives of gauge-invariant objects in $\lcal_{\mathrm SM}$ are
skipped throughout the paper, as they give no physical effects. At the
dimension-five and -six levels, we encounter no gauge-invariant
operators that are built out of non-abelian gauge fields only, and equal to
total derivatives of gauge-variant objects. At the dimension-four level, the
two possible such terms
$\wt G^A_{\mu\nu} G^{A\mu\nu}= 4 \eps^{\mu\nu\rho\sigma} 
  \p_\mu \left( G^A_\nu \p_\rho G^A_\sigma - \f13 g_s f^{ABC} G^A_\nu G^B_\rho G^C_\sigma\right)$~
and the analogous~ $\wt W^I_{\mu\nu}W^{I\mu\nu}$~ should be understood as
implicitly present on the r.h.s of Eq.~(\ref{SMlag4}).  They leave the Feynman
rules and EOMs unaffected, showing up in topological quantum effects
only~\cite{Belavin:1975fg,'tHooft:1976fv,Jackiw:1976pf,Callan:1976je,Baluni:1978rf,Crewther:1979pi}.
\begin{table}[t] 
\centering
\renewcommand{\arraystretch}{1.5}
\btb{||c|c||c|c||c|c||} 
\hline \hline
\multicolumn{2}{||c||}{$X^3$} & 
\multicolumn{2}{|c||}{$\vp^6$~ and~ $\vp^4 D^2$} &
\multicolumn{2}{|c||}{$\psi^2\vp^3$}\\
\hline
$Q_G$                & $f^{ABC} G_\mu^{A\nu} G_\nu^{B\rho} G_\rho^{C\mu} $ &  
$Q_\vp$       & $(\vp^\dag\vp)^3$ &
$Q_{e\vp}$           & $(\vp^\dag \vp)(\bar l_p e_r \vp)$\\
$Q_{\wt G}$          & $f^{ABC} \wt G_\mu^{A\nu} G_\nu^{B\rho} G_\rho^{C\mu} $ &   
$Q_{\vp\Box}$ & $(\vp^\dag \vp)\raisebox{-.5mm}{$\Box$}(\vp^\dag \vp)$ &
$Q_{u\vp}$           & $(\vp^\dag \vp)(\bar q_p u_r \tvp)$\\
$Q_W$                & $\eps^{IJK} W_\mu^{I\nu} W_\nu^{J\rho} W_\rho^{K\mu}$ &    
$Q_{\vp D}$   & $\left(\vp^\dag D^\mu\vp\right)^\star \left(\vp^\dag D_\mu\vp\right)$ &
$Q_{d\vp}$           & $(\vp^\dag \vp)(\bar q_p d_r \vp)$\\
$Q_{\wt W}$          & $\eps^{IJK} \wt W_\mu^{I\nu} W_\nu^{J\rho} W_\rho^{K\mu}$ &&&&\\    
\hline \hline
\multicolumn{2}{||c||}{$X^2\vp^2$} &
\multicolumn{2}{|c||}{$\psi^2 X\vp$} &
\multicolumn{2}{|c||}{$\psi^2\vp^2 D$}\\ 
\hline
$Q_{\vp G}$     & $\vp^\dag \vp\, G^A_{\mu\nu} G^{A\mu\nu}$ & 
$Q_{eW}$               & $(\bar l_p \sigma^{\mu\nu} e_r) \tau^I \vp W_{\mu\nu}^I$ &
$Q_{\vp l}^{(1)}$      & $(\vpj)(\bar l_p \gamma^\mu l_r)$\\
$Q_{\vp\wt G}$         & $\vp^\dag \vp\, \wt G^A_{\mu\nu} G^{A\mu\nu}$ &  
$Q_{eB}$        & $(\bar l_p \sigma^{\mu\nu} e_r) \vp B_{\mu\nu}$ &
$Q_{\vp l}^{(3)}$      & $(\vpjt)(\bar l_p \tau^I \gamma^\mu l_r)$\\
$Q_{\vp W}$     & $\vp^\dag \vp\, W^I_{\mu\nu} W^{I\mu\nu}$ & 
$Q_{uG}$        & $(\bar q_p \sigma^{\mu\nu} T^A u_r) \tvp\, G_{\mu\nu}^A$ &
$Q_{\vp e}$            & $(\vpj)(\bar e_p \gamma^\mu e_r)$\\
$Q_{\vp\wt W}$         & $\vp^\dag \vp\, \wt W^I_{\mu\nu} W^{I\mu\nu}$ &
$Q_{uW}$               & $(\bar q_p \sigma^{\mu\nu} u_r) \tau^I \tvp\, W_{\mu\nu}^I$ &
$Q_{\vp q}^{(1)}$      & $(\vpj)(\bar q_p \gamma^\mu q_r)$\\
$Q_{\vp B}$     & $ \vp^\dag \vp\, B_{\mu\nu} B^{\mu\nu}$ &
$Q_{uB}$        & $(\bar q_p \sigma^{\mu\nu} u_r) \tvp\, B_{\mu\nu}$&
$Q_{\vp q}^{(3)}$      & $(\vpjt)(\bar q_p \tau^I \gamma^\mu q_r)$\\
$Q_{\vp\wt B}$         & $\vp^\dag \vp\, \wt B_{\mu\nu} B^{\mu\nu}$ &
$Q_{dG}$        & $(\bar q_p \sigma^{\mu\nu} T^A d_r) \vp\, G_{\mu\nu}^A$ & 
$Q_{\vp u}$            & $(\vpj)(\bar u_p \gamma^\mu u_r)$\\
$Q_{\vp WB}$     & $ \vp^\dag \tau^I \vp\, W^I_{\mu\nu} B^{\mu\nu}$ &
$Q_{dW}$               & $(\bar q_p \sigma^{\mu\nu} d_r) \tau^I \vp\, W_{\mu\nu}^I$ &
$Q_{\vp d}$            & $(\vpj)(\bar d_p \gamma^\mu d_r)$\\
$Q_{\vp\wt WB}$ & $\vp^\dag \tau^I \vp\, \wt W^I_{\mu\nu} B^{\mu\nu}$ &
$Q_{dB}$        & $(\bar q_p \sigma^{\mu\nu} d_r) \vp\, B_{\mu\nu}$ &
$Q_{\vp u d}$   & $i(\tvp^\dag D_\mu \vp)(\bar u_p \gamma^\mu d_r)$\\
\hline \hline
\etb
\caption{\sf Dimension-six operators other than the four-fermion ones.\label{tab:no4ferm}}
\end{table}

\newsection{ The complete set of dimension-five and -six operators \label{sec:results}}

This Section is devoted to presenting our final results (derived in
  Secs.~\ref{sec:bosons}, \ref{sec:twoferm} and \ref{sec:fourferm}) for the
basis of independent operators $Q_n^{(5)}$ and $Q_n^{(6)}$. Their independence
means that no linear combination of them and their Hermitian conjugates
is EOM-vanishing up to total derivatives.

Imposing the SM gauge symmetry constraints on $Q_n^{(5)}$ leaves out just a
single operator~\cite{Weinberg:1979sa}, up to Hermitian conjugation and
flavour assignments. It reads
\be \label{qnunu}
Q_{\nu\nu} = \eps_{jk} \eps_{mn} \vp^j \vp^m (l^k_p)^T C l^n_r 
~\equiv~ (\tvp^\dag l_p)^T C (\tvp^\dag l_r),
\ee
where $C$ is the charge conjugation matrix.\footnote{ 
  In the Dirac representation $C=i\gamma^2\gamma^0$, with Bjorken and
  Drell\cite{Bjorken:1964XX} phase conventions.}
$Q_{\nu\nu}$ violates the lepton number $L$. After the electroweak symmetry
breaking, it generates neutrino masses and mixings. Neither $\lcal_{\mathrm
  SM}^{(4)}$ nor the dimension-six terms can do the job. Thus, consistency of
the SM (as defined by Eq.~(\ref{Leff}) and Tab.~\ref{tab:matter}) with
observations crucially depends on this dimension-five term.

All the independent dimension-six operators that are allowed by the SM gauge
symmetries are listed in Tabs.~\ref{tab:no4ferm} and \ref{tab:4ferm}. Their
names in the left column of each block should be supplemented with
generation indices of the fermion fields whenever necessary, e.g.,
$Q_{lq}^{(1)} \to Q_{lq}^{(1)prst}$. Dirac indices are always contracted
within the brackets, and not displayed. The same is true for the isospin and
colour indices in the upper part of Tab.~\ref{tab:4ferm}. In the lower-left
block of that table, colour indices are still contracted within the brackets,
while the isospin ones are made explicit. Colour indices are displayed only
for operators that violate the baryon number $B$ (lower-right block of
Tab.~\ref{tab:4ferm}). All the other operators in Tabs.~\ref{tab:no4ferm} and
\ref{tab:4ferm} conserve both $B$ and $L$.
\begin{table}[t]
\centering
\renewcommand{\arraystretch}{1.5}
\begin{tabular}{||c|c||c|c||c|c||}
\hline\hline
\multicolumn{2}{||c||}{$(\bar LL)(\bar LL)$} & 
\multicolumn{2}{|c||}{$(\bar RR)(\bar RR)$} &
\multicolumn{2}{|c||}{$(\bar LL)(\bar RR)$}\\
\hline
$Q_{ll}$        & $(\bar l_p \gamma_\mu l_r)(\bar l_s \gamma^\mu l_t)$ &
$Q_{ee}$               & $(\bar e_p \gamma_\mu e_r)(\bar e_s \gamma^\mu e_t)$ &
$Q_{le}$               & $(\bar l_p \gamma_\mu l_r)(\bar e_s \gamma^\mu e_t)$ \\
$Q_{qq}^{(1)}$  & $(\bar q_p \gamma_\mu q_r)(\bar q_s \gamma^\mu q_t)$ &
$Q_{uu}$        & $(\bar u_p \gamma_\mu u_r)(\bar u_s \gamma^\mu u_t)$ &
$Q_{lu}$               & $(\bar l_p \gamma_\mu l_r)(\bar u_s \gamma^\mu u_t)$ \\
$Q_{qq}^{(3)}$  & $(\bar q_p \gamma_\mu \tau^I q_r)(\bar q_s \gamma^\mu \tau^I q_t)$ &
$Q_{dd}$        & $(\bar d_p \gamma_\mu d_r)(\bar d_s \gamma^\mu d_t)$ &
$Q_{ld}$               & $(\bar l_p \gamma_\mu l_r)(\bar d_s \gamma^\mu d_t)$ \\
$Q_{lq}^{(1)}$                & $(\bar l_p \gamma_\mu l_r)(\bar q_s \gamma^\mu q_t)$ &
$Q_{eu}$                      & $(\bar e_p \gamma_\mu e_r)(\bar u_s \gamma^\mu u_t)$ &
$Q_{qe}$               & $(\bar q_p \gamma_\mu q_r)(\bar e_s \gamma^\mu e_t)$ \\
$Q_{lq}^{(3)}$                & $(\bar l_p \gamma_\mu \tau^I l_r)(\bar q_s \gamma^\mu \tau^I q_t)$ &
$Q_{ed}$                      & $(\bar e_p \gamma_\mu e_r)(\bar d_s\gamma^\mu d_t)$ &
$Q_{qu}^{(1)}$         & $(\bar q_p \gamma_\mu q_r)(\bar u_s \gamma^\mu u_t)$ \\ 
&& 
$Q_{ud}^{(1)}$                & $(\bar u_p \gamma_\mu u_r)(\bar d_s \gamma^\mu d_t)$ &
$Q_{qu}^{(8)}$         & $(\bar q_p \gamma_\mu T^A q_r)(\bar u_s \gamma^\mu T^A u_t)$ \\ 
&& 
$Q_{ud}^{(8)}$                & $(\bar u_p \gamma_\mu T^A u_r)(\bar d_s \gamma^\mu T^A d_t)$ &
$Q_{qd}^{(1)}$ & $(\bar q_p \gamma_\mu q_r)(\bar d_s \gamma^\mu d_t)$ \\
&&&&
$Q_{qd}^{(8)}$ & $(\bar q_p \gamma_\mu T^A q_r)(\bar d_s \gamma^\mu T^A d_t)$\\
\hline\hline
\multicolumn{2}{||c||}{$(\bar LR)(\bar RL)$ and $(\bar LR)(\bar LR)$} &
\multicolumn{4}{|c||}{$B$-violating}\\\hline
$Q_{ledq}$ & $(\bar l_p^j e_r)(\bar d_s q_t^j)$ &
$Q_{duq}$ & \multicolumn{3}{|c||}{$\eps^{\alpha\beta\gamma} \eps_{jk} 
 \left[ (d^\alpha_p)^T C u^\beta_r \right]\left[(q^{\gamma j}_s)^T C l^k_t\right]$}\\
$Q_{quqd}^{(1)}$ & $(\bar q_p^j u_r) \eps_{jk} (\bar q_s^k d_t)$ &
$Q_{qqu}$ & \multicolumn{3}{|c||}{$\eps^{\alpha\beta\gamma} \eps_{jk} 
  \left[ (q^{\alpha j}_p)^T C q^{\beta k}_r \right]\left[(u^\gamma_s)^T C e_t\right]$}\\
$Q_{quqd}^{(8)}$ & $(\bar q_p^j T^A u_r) \eps_{jk} (\bar q_s^k T^A d_t)$ &
$Q_{qqq}$ & \multicolumn{3}{|c||}{$\eps^{\alpha\beta\gamma} \eps_{jn} \eps_{km} 
  \left[ (q^{\alpha j}_p)^T C q^{\beta k}_r \right]\left[(q^{\gamma m}_s)^T C l^n_t\right]$}\\
$Q_{lequ}^{(1)}$ & $(\bar l_p^j e_r) \eps_{jk} (\bar q_s^k u_t)$ &
$Q_{duu}$ & \multicolumn{3}{|c||}{$\eps^{\alpha\beta\gamma} 
  \left[ (d^\alpha_p)^T C u^\beta_r \right]\left[(u^\gamma_s)^T C e_t\right]$}\\
$Q_{lequ}^{(3)}$ & $(\bar l_p^j \sigma_{\mu\nu} e_r) \eps_{jk} (\bar q_s^k \sigma^{\mu\nu} u_t)$ &
& \multicolumn{3}{|c||}{}\\
\hline\hline
\end{tabular}
\caption{\sf Four-fermion operators. \label{tab:4ferm}}
\end{table}

The bosonic operators (classes $X^3$, $X^2\vp^2$, $\vp^6$ and $\vp^4 D^2$) are
all Hermitian. Those containing\lb $\wt X_{\mu\nu}$ are CP-odd, while the
remaining ones are CP-even. For the operators containing fermions, Hermitian
conjugation is equivalent to transposition of generation indices in each of
the fermionic currents in classes $(\bar LL)(\bar LL)$, $(\bar RR)(\bar RR)$,
$(\bar LL)(\bar RR)$, and $\psi^2 \vp^2 D^2$ (except for $Q_{\vp u d}$).  For
the remaining operators with fermions, Hermitian conjugates are not listed
explicitly.

If CP is defined in the weak eigenstate basis then
$Q\,\raisebox{.3mm}{$-$}\hspace{-3.5mm}{}_{{}_{(+)}}\, Q^\dag$~ 
are CP-odd (-even) for all the fermionic operators. It follows that
CP-violation by any of those operators requires a non-vanishing imaginary part
of the corresponding Wilson coefficient. However, one should remember that
such a CP is not equivalent to the usual (``experimental'') one defined in the
mass eigenstate basis, just because the two bases are related by a {\em
  complex} unitary transformation.

Counting the entries in Tabs.~\ref{tab:no4ferm} and \ref{tab:4ferm}, we find
15 bosonic operators, 19 single-fermionic-current ones, and 25 $B$-conserving
four-fermion ones. In total, there are 15+19+25=59 independent dimension-six
operators, so long as $B$-conservation is imposed.

\newsection{ Comparison with Ref.~\cite{Buchmuller:1985jz} \label{sec:compar}}

Comparing the $B$-conserving operators in Tabs.~\ref{tab:no4ferm} and
\ref{tab:4ferm} with Eqs.~(3.3)--(3.64) of Ref.~\cite{Buchmuller:1985jz}, one
finds that
\bit
\item[{\it (i)}] The only operator missed in Ref.~\cite{Buchmuller:1985jz} is
$Q_{lequ}^{(3)} = (\bar l_p^j \sigma_{\mu\nu} e_r) \eps_{jk} (\bar q_s^k \sigma^{\mu\nu} u_t)$.
This fact has been already noticed in Refs.~\cite{Buchmuller:1987ur,Arzt:1994gp} where 
$(\bar l_p^j u_t^\alpha) \eps_{jk} (\bar q_s^{k\alpha} e_r) = -\f18 Q_{lequ}^{(3)} -\f12 Q_{lequ}^{(1)}$~
was used instead. Phenomenological implications for top quark physics have
been discussed, e.g., in Ref.~\cite{Grzadkowski:1997cj,Grzadkowski:1995te}.
\item[{\it (ii)}] One linear combination of the three $\vp^4 D^2$-class
  operators in Eqs.~(3.28) and (3.44) of Ref.~\cite{Buchmuller:1985jz} must be
  redundant because this class contains two independent operators only. In
  fact, presence of all the three operators contradicts correct
  arguments given in Sec.~3.5 of that paper.
\item[{\it (iii)}] The number of single-fermionic-current operators in
  Ref.~\cite{Buchmuller:1985jz} becomes equal to ours after removing all the
  16 operators with covariant derivatives acting on fermion fields
  (Eqs.~(3.30)--(3.37) and (3.57)--(3.59) there).  As we shall show in
  Sec.~\ref{sec:twoferm}, all such operators are indeed redundant. This fact
  has been already discussed in
  Refs.~\cite{Grzadkowski:2003tf,Fox:2007in,AguilarSaavedra:2008zc} for most of the
  cases. Note that removing those operators helps in eliminating multiple
  assignment of the same operator names in Ref.~\cite{Buchmuller:1985jz}.
\item[{\it (iv)}] Our use of~ $\Dfb_\mu$ instead of $D_\mu$ in class
  $\psi^2\vp^2 D$ does not affect the formal operator counting, but actually
  reduces the number of terms to be considered. The point is that Hermitian
  conjugates of our operators with $\Dfb_\mu$ have an identical form as
  the listed ones, so they do not need to be considered separately. On the
  other hand, using scalar field derivatives with a positive relative sign
  (opposite to that in Eq.~(\ref{Dfb})) would give redundant operators
  only, i.e. linear combinations of the three $\psi^2\vp^3$-class terms,
  EOM-vanishing objects, and total derivatives. This issue has been already
  noticed in Ref.~\cite{AguilarSaavedra:2009mx}.
\item[{\it (v)}] Fierz identities (for anticommuting fermion fields)
  like the following one:
\be \label{LLfierz}
(\bar\psi_L \gamma_\mu \psi_L)(\bar \chi_L \gamma^\mu \chi_L) =
(\bar\psi_L \gamma_\mu \chi_L)(\bar \chi_L \gamma^\mu \psi_L)
\ee
make 5 out of 29 four-fermion operators in Ref.~\cite{Buchmuller:1985jz}
linearly dependent on the others.\lb For instance,
\be \label{ltrip}
(\bar l_p \gamma_\mu \tau^I l_r)(\bar l_s \tau^I \gamma^\mu l_t) =
2 (\bar l_p^j \gamma_\mu l_r^k)(\bar l_s^k \gamma^\mu l_t^j) - Q_{ll}^{prst} ~=~
2 Q_{ll}^{ptsr} - Q_{ll}^{prst},
\ee
where the identity
\be \label{su2fierz}
\tau^I_{jk}\tau^I_{mn}= 2\delta_{jn}\delta_{mk}-\delta_{jk}\delta_{mn}
\ee
and Eq.~(\ref{LLfierz}) have subsequently been
used. Sec.~\ref{sec:fourferm} contains a full description of the four-fermion
operator classification.
\eit
As far as the operator names and their normalization are concerned, our notation
is close but not identical to the one of Ref.~\cite{Buchmuller:1985jz}.
Taking advantage of the need to modify the notation because of redundant
operator removal, we do it in several places where convenience is the only
issue. 

\newpage\noindent
The complete list of nomenclature and normalization changes reads:
\bit
\item[{\it (i)}] Unnecessary rationals are skipped in front of $Q_{\vp G}$,
  $Q_{\vp W}$, $Q_{\vp B}$, $Q_\vp$, $Q_{ll}$, $Q^{(1)}_{qq}$, $Q^{(3)}_{qq}$,
  $Q_{ee}$, $Q_{uu}$ and $Q_{dd}$.
\item[{\it (ii)}] $T^A$ instead of $\lambda^A$ are used in $Q_{uG}$, $Q_{dG}$
  and $Q^{(8)}_{\ldots}$.
\item[{\it (iii)}] Fierz transformation and multiplication by $(-2)$ is
  applied in our $(\bar LL)(\bar RR)$ class to avoid crossed colour and Dirac
  index contractions, and to make the notation somewhat more transparent. In
  addition, colour-Fierz transformations are applied to linear combinations of
  the last four operators of this class.
\item[{\it (iv)}] Operator names are changed in many cases to avoid multiple
  use of the same symbols, indicate the presence of essential fields,
  and make the nomenclature more systematic in the four-fermion sector.  In
  particular, the names are modified for $Q_{\vp WB}$, $Q_{\vp \wt WB}$,
  $Q_{\vp u d}$, as well as in the whole $(\bar LR)(\bar RL)$ and $(\bar
  LR)(\bar LR)$ classes.
\eit
One of the reasons for naming our operators with ``Q'' rather than with ``O''
is to indicate that many notational details have changed.  As far as
Sec.~\ref{sec:not} is concerned, we have followed
Ref.~\cite{Buchmuller:1985jz} everywhere except for sign conventions for the
Yukawa couplings in Eq.~(\ref{SMlag4}) and inside covariant derivatives~(Eq.~(\ref{D-sign})).
The latter affects signs of operators in classes $X^3$ and $\psi^2 X\vp$.

\newsection{Bosonic operator classification \label{sec:bosons}}

Building blocks for the SM Lagrangian are the matter fields from
Tab.~\ref{tab:matter}, the field strength tensors $X_{\mu\nu} \in 
\{G^A_{\mu\nu}, W^I_{\mu\nu}, B_{\mu\nu}\}$ and covariant derivatives of
those objects.\footnote{
  If the requirement of gauge invariance was relaxed, gauge fields and their
  fully symmetrized derivatives like $\p_{(\mu_1}\ldots\p_{\mu_n}G_{\nu)}^A$
  would be the only additional objects.  No expression depending on such 
    terms could be gauge-invariant because one can simultaneously nullify
  all of them at any given spacetime point by an appropriate gauge
  transformation.}
Using them and imposing just the global~ $SU(3)_C\times SU(2)_L\times
U(1)_Y$~ symmetry is sufficient to find all the gauge-invariant operators
in $\lcal_{\mathrm SM}$.

Purely bosonic operators must contain an even number of the Higgs
fields $\vp$ (because of the $SU(2)_L$ representation tensor product
constraints), and an even number of covariant derivatives $D$ (because all the
Lorentz indices must be contracted).  Since both $\vp$ and $D$ have
canonical dimension one, while $X$ has dimension two, no dimension-five
operators can arise in the purely bosonic sector. The only possibilities for
the dimension-six bosonic operator field contents are thus $X^3$, $X^2\vp^2$,
$X^2 D^2$, $X\vp^4$, $X D^4$, $X \vp^2 D^2$, $\vp^6$, $\vp^4 D^2$ and $\vp^2
D^4$.

The class $X \vp^4$ is empty because of the antisymmetry of $X$ and absence of
any other objects with Lorentz indices to be contracted. We can also skip $X
D^4$ because all the possible contractions (including those with
$\eps_{\mu\nu\rho\sigma}$) lead to appearance of at least one covariant
derivative commutator $[D_\mu,D_\nu]\sim X_{\mu\nu}$, which moves us to the
$X^2 D^2$ class.

In the following, we shall show that all the possible operators in classes
$\vp^2 D^4$, $\vp^2 X D^2$ and $X^2 D^2$ reduce by the EOMs either to
operators containing fermions or to classes $X^3$, $X^2\vp^2$, $\vp^6$ and
$\vp^4 D^2$. Next, we shall verify that representatives of the latter four
classes in Tab.~\ref{tab:no4ferm} indeed form a complete set of bosonic
operators. 

Since the necessary classical EOMs are going to be used at the
$\ocal\left(\f{1}{\Lambda^2}\right)$ level, and we are not interested in
$\ocal\left(\f{1}{\Lambda^3}\right)$ effects, we can neglect all the 
$\ocal\left(\f{1}{\Lambda}\right)$ terms in the EOMs, i.e. derive them
from $\lcal_{\mathrm SM}^{(4)}$ alone. We get then 
\bea
\left( D^\mu D_\mu \vp \right)^j &=& m^2 \vp^j ~-~ \lambda \left( \vp^\dag \vp \right) \vp^j ~-~ 
\bar e\, \Gamma_{\!e}^\dag l^j ~+~ 
\eps_{jk} \bar{q}^k\, \Gamma_{\!u} u ~-~ 
\bar d\, \Gamma_{\!d}^\dag q^j,\nnb\\[2mm]
\left(D^\rho G_{\rho\mu}\right)^A &=& g_s \left(\bar q \gamma_\mu T^A q
~+~ \bar u \gamma_\mu T^A u ~+~ \bar d \gamma_\mu T^A d\right),\nnb\\[2mm]
\left(D^\rho W_{\rho\mu}\right)^I &=& \f{g}{2}\left( \vpjt ~+~ \bar l \gamma_\mu \tau^I l
                                               ~+~ \bar q \gamma_\mu \tau^I q\right),\nnb\\[2mm]
\p^\rho B_{\rho\mu} &=& g'Y_\vp\,\vpj + g' \sum_{\psi \in \{l,e,q,u,d\}} Y_\psi\,\bar\psi\gamma_\mu\psi.
\label{EOMbos}
\eea
Our ordering of operator classes is such that those containing fewer
  covariant derivatives are considered to be ``lower''. Throughout the paper,
  operators are going to be reduced from higher to lower classes. For classes
  containing equal numbers of derivatives, ordering is defined by the
  number of $X$ tensors, i.e. lower classes contain fewer $X$ tensors.\\[3mm]
\raisebox{1mm}{$\boxed{\vp^2 D^4}$}~ 
In this class, we can restrict our attention to operators where all the
derivatives act on a single $\vp$ field, because other possibilities are
equivalent to them up to total derivatives. Contractions with
$\eps_{\mu\nu\rho\sigma}$ can be ignored because they lead to appearance of
$[D_\mu,D_\nu]\sim X_{\mu\nu}$, which moves us to lower classes
containing $X$. For the same reason, ordering of the covariant derivatives
acting on $\vp$ can be chosen at will. We use this freedom to get $D^\mu D_\mu
\vp$ as a part of each of the considered operators. This moves us by the EOM
to lower classes~ $\vp^4 D^2$,~ $\psi^2 \vp D^2$,~ and 
dimension-four operators multiplied by $m^2$.\\[3mm]
\raisebox{1mm}{$\boxed{\vp^2 X D^2}$}~ 
Here, we allow for $X$ being possibly dual, and forget about
$\eps_{\mu\nu\rho\sigma}$ otherwise. Indices of $X$ cannot be contracted with
themselves, so they need to be contracted with both derivatives. We need to
consider three cases: {\it (i)} Each of the derivatives acts on a different
$\vp$. We can eliminate this possibility ``by parts'', ignoring total
derivatives. {\it (ii)} Both derivatives act on a single object. We obtain
$[D_\mu,D_\nu]\sim X_{\mu\nu}$ and get moved to the $\vp^2 X^2$ class.  {\it
  (iii)} One of the derivatives acts on $X$, and one on $\vp$. We can take
  advantage either of the gauge field EOM (for the usual
tensor) or of the Bianchi identity $D^\rho \wt X_{\rho\mu}=0$ (for the
dual tensor). The EOM moves us to
lower classes~ $\vp^4 D^2$~ and~ $\psi^2 \vp^2 D$.\\[3mm]
\raisebox{1mm}{$\boxed{X^2 D^2}$}~ 
Similarly to the $\vp^2 D^4$ case, we can restrict our attention to operators
where all the derivatives act on a single tensor. If both derivatives are
contracted with $\eps_{\mu\nu\rho\sigma}$ or with a single tensor, we
obtain $[D_\mu,D_\nu]\sim X_{\mu\nu}$, and get moved to the $X^3$ class. Other
contractions with $\eps_{\mu\nu\rho\sigma}$ produce dual tensors. Thus, we
allow the non-differentiated tensor to be possibly dual, and forget about
$\eps_{\mu\nu\rho\sigma}$ otherwise. If each of the derivatives is contracted
with a different tensor, we can use $[D_\mu,D_\nu]\sim X_{\mu\nu}$ to choose
their ordering in such a way that $D^\rho X_{\rho\mu}$ arises. In consequence,
the operator gets reduced by the EOM to lower classes~ $\vp^2 X D^2$~
and~ $\psi^2 X D$. 

The last possibility to consider is when the two
derivatives are contracted with themselves:
\be
{}^{{}^{{}^(}}\!\! \wt X\! {}^{{}^{\,{}^)\!}} {}^{\mu\nu} 
D^\rho D_\rho X_{\mu\nu} = -
{}^{{}^{{}^(}}\!\! \wt X\! {}^{{}^{\,{}^)\!}} {}^{\mu\nu} 
\left( D^\rho D_\mu X_{\nu\rho} + D^\rho D_\nu X_{\rho\mu}\right)
= \boxed{X^3}+\boxed{\vp^2 X D^2} + \boxed{\psi^2 X D} + \boxed{E}\, ,
\ee
In the first step, the Bianchi identity $D_{[\rho} X_{\mu\nu]} = 0$ has been
used. Next, $[D_\rho,D_\alpha]\sim X_{\rho\alpha}$ followed by the EOM for $X$
have been applied. The symbol $\boxed{E}$ stands for EOM-vanishing
operators.\\[2mm]
\raisebox{1mm}{$\boxed{X^3}$}~ 
Here we begin to encounter classes whose representatives do appear in
Tab.~\ref{tab:no4ferm}. To indicate that the tensors may be different, we
denote them by $X$, $Y$ and $Z$ in this paragraph. Allowing one of them to
be dual, we can forget about $\eps_{\mu\nu\rho\sigma}$ otherwise. The only
non-vanishing and independent contraction of Lorentz indices reads~
$X_\mu^{~\nu\,} Y_\nu^{~\rho} Z_\rho^{~\mu}$.~
This implies that all the three tensors {\em must} be different, because~
$X_{\alpha\mu} X_{\beta\nu} Z^{\mu\nu} g^{\alpha\beta}=0$~
by the antisymmetry of $Z$. Moreover, neither of the two tensors can be related by duality because~
$X_\mu^{~\nu} \wt X_\nu^{~\rho} = -\f14 \delta_\mu^\rho X_{\alpha\beta} \wt X^{\alpha\beta}$~
is symmetric in the indices $(\mu\rho)$, while $Z$ is antisymmetric. It
follows that (in particular)~
$B_\mu^{~\nu} W_\nu^{I\rho} \wt W_\rho^{I\mu}=0$,~
i.e. symmetric singlets in products of two adjoint representations are absent
in the considered operator class. The only other option to get a gauge singlet
from three different tensors is to use the structure constants
$f^{ABC}$ or $\eps^{IJK}$. This leads us to a conclusion that the four $X^3$-class 
operators listed in Tab.~\ref{tab:no4ferm} are indeed the only possibilities.\\[2mm]
\raisebox{1mm}{$\boxed{X^2\vp^2}$}~
The Higgs field products combine to singlets or triplets of $SU(2)_L$.
Hypercharge constraints imply that they must be of the form~ $\vp^\dag \vp$~
or~ $\vp^\dag \tau^I \vp$~(but not, e.g., $\vp^\dag \tau^I \tvp$). The eight
$X^2\vp^2$-class operators in Tab.~\ref{tab:no4ferm} contain all the possible
contractions of two field-strength tensors that form singlets or triplets of
$SU(2)_L$, and singlets of $SU(3)_C$.\\[2mm]
\raisebox{1mm}{$\boxed{\vp^6}$}~
For the total hypercharge to vanish, exactly three of the Higgs fields must be
complex conjugated. Grouping the six fields into $\vp^\star \vp$ pairs, and writing
them as in the previous case, we are led to consider tensor products of
singlets and triplets of $SU(2)_L$. Three triplets can combine to an overall
singlet only in a fully antisymmetric manner, which gives zero in our case
because all the triplets are identical
($\eps^{IJK} (\vp^\dag \tau^I \vp)(\vp^\dag \tau^J \vp)(\vp^\dag \tau^K \vp)=0$).~
Two triplets and one singlet combine to an overall singlet as 
$(\vp^\dag \tau^I \vp)(\vp^\dag \tau^I \vp)(\vp^\dag \vp)$ 
that equals to $(\vp^\dag \vp)^3$ thanks to Eq.~(\ref{su2fierz}). Thus, the
only independent operator in the considered class is the very $(\vp^\dag \vp)^3$.\\[2mm]
\raisebox{1mm}{$\boxed{\vp^4 D^2}$}~ 
Hypercharge constraints imply that exactly two $\vp$ fields must be
complex-conjugated. Since the two derivatives must be contracted, either
  they act on two different $\vp$ fields, or the EOM moves the operator to
  lower classes. If they act on two conjugated or two unconjugated fields,
we eliminate those possibilities ``by parts''. If one of them acts on a
conjugated field, and the other on an unconjugated one, our $SU(2)_L$ tensor
product contains four distinct fundamental representations, which means that
exactly two independent singlets must be present. Below, we write them on the
l.h.s.\ as products of triplets and singlets, while the r.h.s.\ explains (via
the Leibniz rule) what combinations give the two simple $\vp^4 D^2$-class
operators in Tab.~\ref{tab:no4ferm}:
\mathindent0cm
\bea
(\vp^\dag\tau^I \vp)\left[(D_\mu\vp)^\dag\tau^I (D^\mu \vp)\right]
&\stackrel{(4.3)}{=}& 
2 \left(\vp^\dag D^\mu\vp\right)^\star \left(\vp^\dag D_\mu\vp\right) -
(\vp^\dag \vp)\left[(D_\mu\vp)^\dag (D^\mu \vp)\right],\nnb\\[2mm]
(\vp^\dag \vp)\left[(D_\mu\vp)^\dag (D^\mu \vp)\right] &\stackrel{(5.1)}{=}&  
\f12 (\vp^\dag \vp) \raisebox{-.5mm}{$\Box$}(\vp^\dag \vp) ~+~ \boxed{\psi^2\vp^3} ~+~
\boxed{\vp^6} ~+~ m^2\, \boxed{\vp^4} ~+~ \boxed{E}\, .
\eea
\mathindent1cm

\newsection{ Single-fermionic-current operator classification \label{sec:twoferm}}

To make general arguments simple, it is convenient to think first in terms of
only left-handed fermions $\psi\in \{l,e^c,q,u^c,d^c\}$, i.e. to use charge
conjugates of the $SU(2)_L$-singlet fermions as fundamental fields. In
such a case, we have only three possibilities for fermionic currents (up to
h.c.):~
$\bar\psi_1 \gamma_\mu \psi_2$,~ $\psi_1^T C \psi_2$~ and~ $\psi_1^T C \sigma_{\mu\nu} \psi_2$.~
  Considering bosonic objects with appropriate numbers of Lorentz indices and
  ignoring $X^\mu_{~\mu}=0$, complete sets of building blocks for our operators
  are easily determined for each of the currents. They read\footnote{
Bosonic terms leading to dimension-five and -six operators are collected in separate brackets.}
\be \label{currents}
\baa{rclcl}
\bar\psi_1 \gamma_\mu \psi_2:      &~~~& (\vp D),  &~~~& (X D,~ \vp^2 D,~ D^3), \\[1mm]
\psi_1^T C \psi_2:                  && (\vp^2,~ D^2),  && (\vp^3,~ \vp D^2), \\[1mm]
\psi_1^T C \sigma_{\mu\nu} \psi_2:  && (X,~ D^2),      && (X\vp,~ \vp D^2).
\eaa 
\ee

A brief look into Tab.~\ref{tab:matter} ensures that hypercharges of the
currents involving $C$ never vanish, while hypercharges of the vector currents
never equal $\pm 1/2$. Consequently, classes $\psi^2 X$,~ $\psi^2 D^2$ and~
$\psi^2\vp D$~ are empty. Moreover, the Higgs field products in class
$\psi^2\vp^2$ must give non-zero hypercharges, in which case the only
possibilities are $\pm 1$. There is only a single fermionic current that can
compensate such a hypercharge, namely the one built out of two lepton
doublets. Thus, we obtain the field content of the operator in
Eq.~(\ref{qnunu}). The isospin structure of that operator is the only
available one given the antisymmetry of $\eps_{jk}$ and the presence of just a
single Higgs doublet in the SM. This completes our discussion of
dimension-five operators.

In the dimension-six case, the number of Higgs fields associated with scalar
and tensor fermionic currents is always odd. Consequently, those currents must
form isospin doublets. In the standard notation with right-handed singlets,
they read $\bar\psi_1 \psi_2$ and $\bar\psi_1 \sigma_{\mu\nu} \psi_2$.
Similarly, vector currents can only form isospin singlets or triplets, as
they combine with even numbers of the Higgs fields. Therefore, even if the
isospin singlets are taken right-handed, no vector currents with $C$ enter
into our considerations. We shall thus return to the standard notation in what
follows.

Classical EOMs for the quarks and leptons that we are going to use below read
\mathindent0cm
\be \label{EOMferm} 
\baa{cccccccccc}
i\D l = \Gamma_{\!e} e \vp, &&
i\D e = \Gamma_{\!e}^{\dag} \vp^{\dag} l, &&
i\D q = \Gamma_{\!u} u \tvp + \Gamma_{\!d} d \vp, &&
i\D u = \Gamma_{\!u}^{\dag} \tvp^\dag q, &&
i\D d = \Gamma_{\!d}^{\dag} \vp^\dag q.\! 
\eaa
\ee
\mathindent1cm
Apart from them, two simple Dirac-algebra identities need to be recalled, namely
\be \label{gg} 
\gamma_\mu \gamma_\nu = g_{\mu\nu} - i\sigma_{\mu\nu}\, , \hspace{2cm}
\gamma_\mu \gamma_\nu \gamma_\rho = g_{\mu\nu}\gamma_\rho +g_{\nu\rho}\gamma_\mu 
- g_{\mu\rho}\gamma_\nu - i\eps_{\mu\nu\rho\sigma}\gamma^\sigma\gamma_5\, . 
\ee

Let us now discuss all the dimension-six classes one-by-one.\footnote{
  There are six of them. Note that both the scalar and tensor currents occur in
  the $\psi^2\vp D^2$ case.}\\[3mm]
\raisebox{1mm}{$\boxed{\psi^2 D^3}$}~
Three covariant derivatives are contracted here with a certain $\bar\psi
\gamma_\mu \psi$ current. Similarly as in the previously discussed
classes $\vp^2 D^4$ and $X^2 D^2$, we can remove derivatives acting on
$\bar\psi$\lb ``by parts'', and choose ordering of the derivatives acting on
$\psi$ at will. Choosing the ordering as in $\bar\psi D_\mu D^\mu \D \psi$, we
get an operator that reduces
by the EOMs to class $\psi^2 \vp D^2$.\\[3mm]
\raisebox{1mm}{$\boxed{\psi^2 \vp D^2}$}~ As follows from
  Eq.~(\ref{currents}), this class involves scalar and tensor fermion
  currents only. We remove the derivatives acting on $\bar\psi$ ``by parts'',
and take into account that
$\bar\psi \sigma^{\mu\nu} \psi\, D_\mu D_\nu \vp$~ and
$\vp\bar\psi \sigma^{\mu\nu} D_\mu D_\nu \psi$~ 
belong actually to class~ $\psi^2 X \vp$~ because~ $[D_\mu,D_\nu]\sim X_{\mu\nu}$.
The four remaining possibilities 
[\,$\bar\psi \psi\, D_\mu D^\mu \vp$,~ 
$\vp \bar\psi D_\mu D^\mu \psi$,~ 
$(D_\mu\vp) \bar\psi \sigma^{\mu\nu} D_\nu \psi$~ 
and~ $(D^\mu\vp) \bar\psi D_\mu \psi$\,]
are EOM-reduced to lower classes as follows:
\mathindent0cm
\bea
\bar\psi \psi D_\mu D^\mu \vp &\stackrel{(5.1)}{=}& 
\boxed{\psi^4} + \boxed{\psi^2\vp^3} +  m^2\,\boxed{\psi^2\vp} + \boxed{E}\, ,\nnb\\[1mm]
\vp \bar\psi D_\mu D^\mu \psi &\stackrel{(6.3)}{=}& 
\vp \bar\psi \D \D \psi + \boxed{\psi^2 X \vp}~\stackrel{(6.2)}{=}~
\boxed{\psi^2 X \vp} + \boxed{\psi^2\vp^2 D} + \boxed{E}\, ,\nnb\\[1mm]
(D_\mu\vp) \bar\psi \sigma^{\mu\nu} D_\nu \psi &=&
\f{i}{2} (D_\mu\vp) \bar\psi \left( \gamma^\mu \D - \D \gamma^\mu\right) \psi ~=~
i (D_\mu\vp) \bar\psi \gamma^\mu \D \psi - i (D^\mu\vp) \bar\psi D_\mu \psi\nnb\\[1mm] 
&\stackrel{(6.2)}{=}&
- i (D^\mu\vp) \bar\psi D_\mu \psi + \boxed{\psi^2\vp^2 D} + \boxed{E}\, ,\nnb\\[1mm]
2 (D^\mu\vp) \bar\psi D_\mu \psi &=&
(D^\mu\vp) \bar\psi (\gamma_\mu \D + \D \gamma_\mu ) \psi\nnb\\[1mm]
&=& (D^\mu\vp) \bar\psi \gamma_\mu \D \psi 
- \bar\psi \not\hspace{-1mm}\Db \gamma_\mu \psi\, D^\mu\vp
- \bar\psi \gamma^\nu \gamma^\mu \psi\, D_\nu D_\mu \vp + \boxed{T}\nnb\\[1mm]
&\stackrel{(6.2)}{=}& \boxed{\psi^2\vp^2 D} + \boxed{\psi^4} +\boxed{\psi^2\vp^3} 
+ m^2\,\boxed{\psi^2\vp} +\boxed{\psi^2 X \vp} + \boxed{E} + \boxed{T}\, ,
\eea
\mathindent1cm
where $\boxed{T}$ stands for a total derivative. In the last step above, one should
realize that
\mathindent0cm
\be
\bar\psi \gamma^\nu \gamma^\mu \psi\, D_\nu D_\mu \vp 
\stackrel{(6.3)}{=}
\bar\psi \psi\, D_\mu D^\mu \vp - i \bar\psi \sigma^{\nu\mu} \psi\, D_\nu D_\mu \vp 
\stackrel{(5.1)}{=} \boxed{\psi^4} + \boxed{\psi^2\vp^3} + m^2\,\boxed{\psi^2\vp} 
+ \boxed{\psi^2 X \vp} + \boxed{E}.
\ee
\mathindent1cm
\raisebox{1mm}{$\boxed{\psi^2 X D}$}~
As in several previous cases, we allow for $X$ being possibly dual, and forget
about $\eps_{\mu\nu\rho\sigma}$ otherwise. Since we deal here with
$\bar\psi\gamma_\mu \psi$ currents only, the derivative must be contracted
with $X$. If it acts on $X$, we obtain either the gauge field EOM (for
the usual tensor) or the Bianchi identity $D^\rho \wt X_{\rho\mu}=0$ (for the
dual tensor). The EOM moves us to lower classes ${\psi^2 \vp^2 D}$ and
$\psi^4$. Removing ``by parts'' terms with derivatives acting on $\bar\psi$,
we find that the only expression still to be considered is $X^{\mu\nu}
\bar\psi \gamma_\mu D_\nu \psi$. It gets reduced to lower classes as follows:
\mathindent0cm
\bea
X^{\mu\nu} \bar\psi \gamma_\mu D_\nu \psi &=& \f12 X^{\mu\nu} \bar\psi (\gamma_\mu \gamma_\nu \D 
+ \gamma_\mu \D \gamma_\nu) \psi ~=~ \f12 X^{\mu\nu} \bar\psi (\gamma_\mu \gamma_\nu \D 
- \D \gamma_\mu \gamma_\nu) \psi + X^{\mu\nu} \bar\psi \gamma_\nu D_\mu \psi\nnb\\[1mm]
&\stackrel{(*)}{=}& 
\f14 X^{\mu\nu} \bar\psi (\gamma_\mu \gamma_\nu \D - \D \gamma_\mu \gamma_\nu) \psi ~=~
\f14 X^{\mu\nu} \bar\psi \gamma_\mu \gamma_\nu \D \psi + 
\f14 \bar\psi \not\hspace{-1mm}\Db \gamma_\mu \gamma_\nu \psi\, X^{\mu\nu}\nnb\\[1mm] 
&+& \f14 \bar\psi \gamma_\rho \gamma_\mu \gamma_\nu \psi\, D^\rho X^{\mu\nu} 
+\,\boxed{T} ~\stackrel{(6.2)}{=}~
\boxed{\psi^2 X \vp} + \boxed{\psi^2\vp^2 D} + \boxed{\psi^4} + \boxed{E} + \boxed{T}\, .
\eea
\mathindent1cm
In the third step above (denoted by $(*)$), we have taken into account that the last
term in the preceding expression is equal to our initial operator but with an
opposite sign. In the last step, we have used the equality
\mathindent0cm
\be \label{gggX}
\bar\psi \gamma_\rho \gamma_\mu \gamma_\nu \psi\,D^\rho X^{\mu\nu}~\stackrel{(6.3)}{=}~
2\, \bar\psi \gamma^\mu \psi \, D^\rho X_{\rho\mu}
- i\eps_{\rho\mu\nu\sigma}\, \bar\psi \gamma^\sigma\gamma_5 \psi\, D^\rho X^{\mu\nu} ~=~ 
\boxed{\psi^2\vp^2 D} + \boxed{\psi^4} + \boxed{E}\, .
\ee
\mathindent1cm
Both the gauge field EOM and the Bianchi identity are necessary in Eq.~(\ref{gggX}),
irrespectively of whether the initial $X$ is dual or not.

\newpage\noindent
\raisebox{1mm}{$\boxed{\psi^2 \vp^3}$}~
According to the arguments given above Eq.~(\ref{EOMferm}), the fermion
current must be an isospin doublet and colour singlet of the form $\bar \psi_1
\psi_2$, i.e. one of those present in the Yukawa terms in
Eq.~(\ref{SMlag4}). The number of conjugated and unconjugated scalar fields in
$\vp^3$ is fixed for each of the fermionic currents by hypercharge
constraints. Combining those scalar fields into an isospin doublet is 
  unique because one of the two doublets in $\hat 2\otimes\hat 2\otimes\hat 2$
  vanishes in each of the cases due to
$\vp^\dag\tvp = \eps_{jk}(\vp^j)^\star(\vp^k)^\star = 0 = \eps_{jk}\vp^j\vp^k$.
Consequently, the only possibilities for this class are the Yukawa
terms multiplied by $\vp^\dag\vp$,~ as
in the upper-right block of Tab.~\ref{tab:no4ferm}.\\[3mm]
\raisebox{1mm}{$\boxed{\psi^2 X \vp}$}~
The antisymmetric tensor and the single Higgs field enforce the fermion
current to be an isospin doublet of the form $\bar \psi_1 \sigma^{\mu\nu}
\psi_2$. Vanishing total hypercharge can be obtained only if the Higgs field
combines with the currents in analogy to the standard Yukawa terms in
Eq.~(\ref{SMlag4}). Couplings with $B_{\mu\nu}$ in Tab.~\ref{tab:no4ferm} show
this analogy most transparently. The tensors $W^I_{\mu\nu}$ and
  $G^A_{\mu\nu}$ need to be contracted with isospin triplets and colour
  octets, respectively, which can be formed just in a single way for each of
  the cases, as in Tab.~\ref{tab:no4ferm}.  Dualizing the $X$ tensor in any of
  the $\psi^2 X \vp$-class operators in that table would not give anything new
  because of the identities
  $\eps_{\alpha\beta\mu\nu}\sigma^{\mu\nu}=2i\sigma_{\alpha\beta}\gamma_5$ and
  $\gamma_5\psi_{L,R}=\mp\psi_{L,R}$.\\[3mm]
\raisebox{1mm}{$\boxed{\psi^2 \vp^2 D}$}~
If the derivative acts on any of the fermion fields, its contraction with the
$\bar\psi\gamma_\mu \psi$ current\lb produces EOMs and moves us to the previously
discussed lower class $\psi^2 \vp^3$. Thus, it is sufficient to consider
derivatives acting on the scalars only. The Higgs fields can form isospin
singlets or triplets, and are colour singlets. The fermion currents must
follow the same selection rules, which allows precisely the currents
listed in the ${\psi^2 \vp^2 D}$-class block of Tab.~\ref{tab:no4ferm}, up to
Hermitian conjugation of the $\bar u\gamma_\mu d$ current. Hypercharge
constraints determine the number of conjugated and unconjugated Higgs fields.
We begin with removing ``by parts'' derivatives acting on one of the scalars,
and forming isospin singlets or triplets from products of $\vp_1$ and
$D_\mu\vp_2$, according to the structure of the corresponding fermion currents,
which gives unique expressions in all the cases. This way we get operators
differing from the ones in Tab.~\ref{tab:no4ferm} only by the presence of $D$
instead of the $\Dfb$.  However, we cannot terminate at this point because the
operators without $\Dfb$ are not Hermitian, and we still need to check whether
their Hermitian conjugates are independent from them or not. Such a question
does not arise for any other block of Tabs.~\ref{tab:no4ferm}
and~\ref{tab:4ferm} because all the other operators are either manifestly
Hermitian (up to flavour permutations in the upper part of
Tab.~\ref{tab:4ferm}) or their Hermitian conjugates are manifestly
independent (due to absence of hypercharge-conjugated fermion pairs). Such a
manifest independence occurs also in the case of $Q_{\vp ud}$ in the
considered class, so we leave it with the usual derivative.\footnote{
Actually,~ $\tvp^\dag \left( D_\mu - \Db_\mu\right) \vp = 2 \tvp^\dag D_\mu \vp$.} 
In the remaining seven cases (which contain hypercharge-neutral currents),
we form combinations with $\Dfb$ as in Tab.~\ref{tab:no4ferm}, and supplement
them with symmetrized combinations of the form
\mathindent0cm
\be
\left[ \vp^\dag ( D_\mu + \Db_\mu ) \vp\right] \bar\psi \gamma^\mu \psi =
\left[ \p_\mu (\vp^\dag \vp) \right] \bar\psi \gamma^\mu \psi =
(\vp^\dag \vp) \bar\psi ( \D +  \not\hspace{-1mm}\Db )\psi +\boxed{T}
= \boxed{\psi^2\vp^3} + \boxed{E} + \boxed{T}.
\ee
\mathindent1cm
Thus, the symmetrized combinations give redundant operators and can be
ignored.  At this point, our classification of all the single-fermionic-current
operators has been completed.

\newsection{ Four-fermion operator classification \label{sec:fourferm}}

Four fermion operators are the most numerous but very easy to classify. As in
the beginning of the previous Section, we think first in terms of only
left-handed fermions $\psi\in \{l,e^c,q,u^c,d^c\}$. Lorentz-singlet
products of the fermionic currents~(\ref{currents}) and their Hermitian
conjugates never give field contents like $\psi\psi\psi\bar\psi$ or
$\psi\bar\psi\bar\psi\bar\psi$. For the remaining options, we search for
zero-hypercharge products without paying attention to whether they can form
isospin or colour singlets. There are several hundreds of cases to be tested,
which is done in less than a second by a simple computer algebra code. Apart
from trivial results giving products of two zero-hypercharge currents, only a
handful of other
possible field contents are found, namely\\
\be \label{exceptional}
(\bar l \bar e^c d^c q),~
(q u^c q d^c),~
(l e^c q u^c),~
(q q q l),~
(d^c u^c u^c e^c),~
(q q \bar u^c \bar e^c),~
(q l \bar u^c \bar d^c), 
\ee
and their Hermitian conjugates. Apparently, none of them can be eliminated
using $SU(2)_L$ or $SU(3)_C$ constraints. The first three are $B$-conserving,
while the remaining four are $B$-violating.

In the cases with two $\psi$ and two $\bar\psi$ fields in
Eq.~(\ref{exceptional}), it is enough to consider only a single pairing of the four
fields into two~ $\bar\psi_L \gamma_\mu \psi_L$~ currents.\footnote{
There$\;$ is$\;\,$ only one $SL(2,\mathbb{C})$ singlet~ in 
$(0,\f12)\otimes(0,\f12)\otimes(\f12,0)\otimes(\f12,0)$,
which shows up in Eq.~(\ref{LLfierz}).\\[-3mm]}
As far as $SU(2)_L$ is concerned, in each case there are two doublet and two
singlet fields, which gives us only one overall singlet.  Finally, there is
only one $SU(3)_C$ singlet in $\hat{\bar 3}\otimes \hat 3$ for the
$B$-conserving operator, and one in $\hat 3\otimes \hat 3 \otimes \hat 3$ for
the $B$-violating ones. Consequently, we get just a single operator for each
of the three considered field contents. They are given by $Q_{ledq}$, 
  $Q_{duq}$ and $Q_{qqu}$ in Tab.~\ref{tab:4ferm} after passing to the
standard notation with right-handed $SU(2)_L$ singlets.

In the remaining cases in Eq.~(\ref{exceptional}), four left-handed $\psi$
fields occur. Once both the scalar and tensor currents from Eq.~(\ref{currents})
are taken into account, only a single pairing of the fields into currents
needs to be considered.\footnote{
There are only two $SL(2,\mathbb{C})$ singlets in 
$(\f12,0)\otimes(\f12,0)\otimes(\f12,0)\otimes(\f12,0)$.}
Alternatively, one can use the Fierz identity
\be \label{Cfierz}
(\psi_{1L}^T C \sigma_{\mu\nu} \psi_{2L})(\psi_{3L}^T C \sigma^{\mu\nu} \psi_{4L}) 
= -4 (\psi_{1L}^T C \psi_{2L})(\psi_{3L}^T C \psi_{4L}) 
- 8 (\psi_{1L}^T C \psi_{4L})(\psi_{3L}^T C \psi_{2L}) 
\ee
to get rid of the tensor currents. We choose the latter option everywhere
except for the $(l e^c q u^c)$ field content ($Q_{lequ}^{(1)\dag}$ and
$Q_{lequ}^{(3)\dag}$), where we want to retain colour index contractions within
the currents. In the three other cases ($(q u^c q d^c)$, $(q q q l)$ and $(d^c
u^c u^c e^c)$), considering two different pairings amounts merely to a
different generation index assignment, because two fields of the same type are
always present. Once the fields are paired into currents, we determine all the
possible isospin and colour index contractions. They are identified as $Q_{quqd}^{(1)\dag}$,
$Q_{quqd}^{(8)\dag}$, $Q_{qqq}$ and $Q_{duu}^\dag$.

This way we have completed establishing a basis for all the operators that
cannot be written as products of zero-hypercharge currents, i.e. classes
$(\bar LR)(\bar RL)$, $(\bar LR)(\bar LR)$ and $B$-violating in
Tab.~\ref{tab:4ferm}. The $B$-violating ones are identical to those in
Ref.~\cite{Abbott:1980zj} where the original classification of
Refs.~\cite{Weinberg:1979sa,Wilczek:1979hc} was corrected.\footnote{
The original version of our paper contained five $B$-violating operators,
as in Eqs.~(1.1)-(1.5) of Ref.~\cite{Abbott:1980zj}. However, only four of
them are linearly independent, as shown in Eqs.~(1.7)-(1.11) of that article
(see also appendix A of Ref.~\cite{Alonso:2014zka}).}
It is worth recalling that $Q_{qqq}$ vanishes in the flavour-diagonal 
case, as follows from Eq.~(1.11) of Ref.~\cite{Abbott:1980zj}.

If the field content of a four-fermion operator allows to write it as a
product of two zero-hypercharge currents, we write it like that using the
Fierz identity~(\ref{LLfierz}) if necessary. Next, we pass to the standard
notation with right-handed $SU(2)_L$ singlets, which splits the considered set
into classes $(\bar LL)(\bar LL)$, $(\bar RR)(\bar RR)$ and $(\bar LL)(\bar
RR)$ in Tab.~\ref{tab:4ferm}. It remains to convince oneself that the
operators listed there indeed form complete bases for those classes. In the
beginning, one should consider all the possible products of currents that form
isospin singlets or triplets, and colour singlets or octets. Next, it is
possible to eliminate several cases in the $(\bar LL)(\bar LL)$ and
$(\bar RR)(\bar RR)$ classes using the relation (\ref{su2fierz}) together with
\be \label{su3fierz}
T^A_{\alpha\beta} T^A_{\kappa\lambda} = \f12 \delta_{\alpha\lambda}
\delta_{\kappa\beta} - \f16 \delta_{\alpha\beta} \delta_{\kappa\lambda}\, ,
\ee
and the Fierz identity~(\ref{LLfierz}) or its right-handed counterpart.  It is
essential to take into account that all the possible flavour assignments
are included in Tab.~\ref{tab:4ferm}. One of such simplifications has been
already shown in Eq. (\ref{ltrip}). The remaining ones read
\mathindent0cm
\bea
&& \nnb\\[-5mm]
(\bar u_p \gamma_\mu T^A u_r)(\bar u_s T^A \gamma^\mu u_t) &\stackrel{(7.3)}{=}&
\f12 (\bar u_p^\alpha \gamma_\mu u_r^\beta)(\bar u_s^\beta \gamma^\mu u_t^\alpha) 
- \f16 Q_{uu}^{prst} ~=~
\f12 Q_{uu}^{ptsr} - \f16 Q_{uu}^{prst},\label{uuFierz}\\[4mm]
(\bar d_p \gamma_\mu T^A d_r)(\bar d_s T^A \gamma^\mu d_t) &\stackrel{(7.3)}{=}&
\f12 (\bar d_p^\alpha \gamma_\mu d_r^\beta)(\bar d_s^\beta \gamma^\mu d_t^\alpha) 
- \f16 Q_{dd}^{prst} ~=~
\f12 Q_{dd}^{ptsr} - \f16 Q_{dd}^{prst},\\[4mm]
(\bar q_p \gamma_\mu T^A q_r)(\bar q_s T^A \gamma^\mu q_t) &\stackrel{(7.3)}{=}&
\f12 (\bar q_p^{\alpha j} \gamma_\mu q_r^{\beta j})(\bar q_s^{\beta k} \gamma^\mu q_t^{\alpha k}) 
- \f16 Q_{qq}^{(1)prst}\nnb\\[2mm] 
&\stackrel{(4.1)}{=}&
\f12 (\bar q_p^{\alpha j} \gamma_\mu q_t^{\alpha k})(\bar q_s^{\beta k} \gamma^\mu q_r^{\beta j}) 
- \f16 Q_{qq}^{(1)prst}\nnb\\[2mm] 
&\stackrel{(4.3)}{=}&
\f14 Q_{qq}^{(3)ptsr} + \f14 Q_{qq}^{(1)ptsr} - \f16 Q_{qq}^{(1)prst},\\[4mm]
(\bar q_p \gamma_\mu T^A \tau^I q_r)(\bar q_s T^A \tau^I \gamma^\mu q_t) &\stackrel{(7.3)}{=}&
\f12 (\bar q_p^\alpha \gamma_\mu \tau^I q_r^\beta)(\bar q_s^\beta \gamma^\mu \tau^I q_t^\alpha) 
- \f16 Q_{qq}^{(3)prst}\nnb\\[2mm]
&\stackrel{(4.3)}{=}&  
 (\bar q_p^{\alpha j} \gamma_\mu q_r^{\beta k})(\bar q_s^{\beta k} \gamma^\mu q_t^{\alpha j}) 
-\f12 (\bar q_p^{\alpha j} \gamma_\mu q_r^{\beta j})(\bar q_s^{\beta k} \gamma^\mu q_t^{\alpha k}) 
- \f16 Q_{qq}^{(3)prst}\nnb\\[2mm] 
&\stackrel{(4.1)}{=}& Q_{qq}^{(1)ptsr} 
-\f12 (\bar q_p^{\alpha j} \gamma_\mu q_t^{\alpha k})(\bar q_s^{\beta k} \gamma^\mu q_r^{\beta j})
- \f16 Q_{qq}^{(3)prst}\nnb\\[2mm] 
&\stackrel{(4.3)}{=}&  
-\f14 Q_{qq}^{(3)ptsr} + \f34 Q_{qq}^{(1)ptsr} - \f16 Q_{qq}^{(3)prst}.\\[-3mm] \label{qq81Fierz}
&& \nnb
\eea
\mathindent1cm
Establishing the above relations completes the proof that our four-fermion
operator set in Tab.~\ref{tab:4ferm} is indeed exhaustive.

\newpage
\newsection{ Conclusions \label{sec:concl}}

A tremendous simplification of the operator basis by the EOMs can be
appreciated by comparing our Tab.~\ref{tab:no4ferm} that contains 34 entries
with Ref.~\cite{Leung:1984ni} where 106 operators involving bosons are present
because no EOM-reduction has been applied. Going down from 106 to 51 with the
help of EOMs in Ref.~\cite{Buchmuller:1985jz} has been a partial success. It
is really amazing that no author of almost 600 papers that quoted
Ref.~\cite{Buchmuller:1985jz} over 24 years has ever decided to rederive the
operator basis from the outset to check its correctness. As the current work
shows, the exercise has been straightforward enough for an
M.$\,$Sc.$\;$thesis~\cite{Iskrzynski:2010xxx,Iskrzynski:2010yyy}. It has
required no extra experience with respect to what was standard already in the
1980's.

From the phenomenological standpoint, it is hard to overestimate the
importance of knowing the explicit form of power-suppressed terms in the SM
Lagrangian. Although their overall number is sizeable, usually very few of
them contribute to a given process. For instance, anomalous $Wtb$ couplings
that can be well tested at the LHC are described by four operators only
($Q_{uW}$, $Q_{dW}$, $Q_{\vp q}^{(3)}$ and $Q_{\vp
  ud}$)~\cite{AguilarSaavedra:2008zc,AguilarSaavedra:2009mx,Grzadkowski:2008mf}.
Given 14 operators in the dimension-four Lagrangian~(\ref{SMlag4}), it is
  actually quite surprising that no more than 59 operators arise at the
  dimension-six level.

It is interesting to note that if the underlying beyond-SM model is a weakly
coupled (perturbative) gauge theory, operators containing field-strength
tensors in Tab.~\ref{tab:no4ferm} cannot be tree-level
generated~\cite{Arzt:1994gp}. In consequence, their Wilson coefficients $C_k$
are typically $\ocal\left(\f{1}{16\pi^2}\right)$. Thus, so long as we are
interested in operators with $\ocal(1)$ coefficients only, as little as 14
entries of Tab.~\ref{tab:no4ferm} remain relevant. Investigations
involving those operators can be found, e.g., in
Refs.~\cite{Grzadkowski:1997cj,Grzadkowski:1995te,Grzadkowski:1995hi}.

\section*{Note Added}

While this article was being completed, a new
  paper~\cite{AguilarSaavedra:2010zi} on four-fermion operator classification
  appeared on the arXiv. The number of independent $B$-conserving operators
  found there is the same as in our Tab.~\ref{tab:4ferm}. The key point are
the identities (\ref{uuFierz})--(\ref{qq81Fierz}) that have not remained
unnoticed~\cite{Hioki:2008XXX}, but we are not aware of mentioning them in the
literature previously in the context of correcting
Ref.~\cite{Buchmuller:1985jz}.

\section*{Acknowledgments}
We would like to thank Wilfried Buchm\"uller and Daniel Wyler for
  correspondence and discussions in years 2008--2010 concerning preparations
  to reanalyzing their operator classification~\cite{Buchmuller:1985jz}.  We
  are grateful to Pawe{\l} Nurowski for helpful advice. The work of
B.G. and M.M. has been supported in part by the Ministry of Science and Higher
Education (Poland) as research project N~N202~006334 (2008-11).  B.G. and
  J.R. acknowledge support of the European Community within the Marie Curie
Research \& Training Networks: ``HEPTOOLS" (MRTN-CT-2006-035505), and
``UniverseNet" (MRTN-CT-2006-035863).  M.M. acknowledges support from
the EU-RTN Programme ``FLAVIAnet'' (MRTN-CT-2006-035482) and from the DFG
through the ``Mercator'' guest professorship programme.  The work of J.R
  has been supported in part by the Ministry of Science and Higher Education
  (Poland) as research projects N~N202~230337 (2009-12) and N~N202~103838 (2010-12).

\setlength {\baselineskip}{0.2in}
 

\begin{thebibliography}{99}

\bibitem{Nakamura:2010XXX}
K. Nakamura {\it et al.} (Particle Data Group)
J.\ Phys.\ G {\bf 37} (2010) 075021, 
{\tt http://pdg.lbl.gov}~.
%
\bibitem{Appelquist:1974tg}
  T.~Appelquist and J.~Carazzone,
  Phys.\ Rev.\  D {\bf 11} (1975) 2856. 
%
\bibitem{Buchmuller:1985jz}
  W.~Buchm\"uller and D.~Wyler,
  Nucl.\ Phys.\  B {\bf 268} (1986) 621. 
%
\bibitem{Politzer:1980me}
  H.~D.~Politzer,
  Nucl.\ Phys.\  B {\bf 172} (1980) 349.~ 
%
\bibitem{KlubergStern:1975hc}
  H.~Kluberg-Stern and J.~B.~Zuber,
  Phys.\ Rev.\  D {\bf 12} (1975) 3159.
%
\bibitem{GrosseKnetter:1993td}
  C.~Grosse-Knetter,
  Phys.\ Rev.\  D {\bf 49} (1994) 6709 
  [hep-ph/9306321].
%
\bibitem{Arzt:1993gz}
  C.~Arzt,
  Phys.\ Lett.\  B {\bf 342} (1995) 189 
  [hep-ph/9304230].
%
\bibitem{Simma:1993ky}
  H.~Simma,
  Z.\ Phys.\  C {\bf 61} (1994) 67 
  [hep-ph/9307274].
%
\bibitem{Wudka:1994ny}
  J.~Wudka,
  Int.\ J.\ Mod.\ Phys.\  A {\bf 9} (1994) 2301
  [hep-ph/9406205].
%
\bibitem{Grzadkowski:2003tf}
  B.~Grzadkowski, Z.~Hioki, K.~Ohkuma and J.~Wudka,
  Nucl.\ Phys.\  B {\bf 689} (2004) 108 
  [hep-ph/0310159].
%
\bibitem{Fox:2007in}
  P.~J.~Fox {\it et al.}, 
  Phys.\ Rev.\  D {\bf 78} (2008) 054008
  [arXiv:0704.1482].
%
\bibitem{AguilarSaavedra:2008zc}
  J.~A.~Aguilar-Saavedra,
  Nucl.\ Phys.\  B {\bf 812} (2009) 181
  [arXiv:0811.3842].
%
\bibitem{AguilarSaavedra:2009mx}
  J.~A.~Aguilar-Saavedra,
  Nucl.\ Phys.\  B {\bf 821} (2009) 215
  [arXiv:0904.2387].
%
\bibitem{Belavin:1975fg} 
  A.~A.~Belavin, A.~M.~Polyakov, A.~S.~Schwartz and Yu.~S.~Tyupkin,
  Phys.\ Lett.\  B {\bf 59} (1975) 85.
%
\bibitem{'tHooft:1976fv}
  G.~'t Hooft,
  Phys.\ Rev.\  D {\bf 14} (1976) 3432
  [Erratum-ibid.\  D {\bf 18} (1978) 2199].
%
\bibitem{Jackiw:1976pf}
  R.~Jackiw and C.~Rebbi,
  Phys.\ Rev.\ Lett.\  {\bf 37} (1976) 172.
%
\bibitem{Callan:1976je}
  C.~G.~.~Callan, R.~F.~Dashen and D.~J.~Gross,
  Phys.\ Lett.\  B {\bf 63} (1976) 334.
%
\bibitem{Baluni:1978rf}
  V.~Baluni,
  Phys.\ Rev.\  D {\bf 19} (1979) 2227.
%
\bibitem{Crewther:1979pi}
  R.~J.~Crewther, P.~Di Vecchia, G.~Veneziano and E.~Witten,
  Phys.\ Lett.\  B {\bf 88} (1979) 123
  [Erratum-ibid.\  B {\bf 91} (1980) 487].
%
\bibitem{Weinberg:1979sa}
  S.~Weinberg,
  Phys.\ Rev.\ Lett.\  {\bf 43} (1979) 1566.
%
\bibitem{Bjorken:1964XX}
  J.~D.~Bjorken and S.~D.~Drell,
  {\it ``Relativistic Quantum Mechanics''},
  McGraw-Hill Inc., 1964.
%
\bibitem{Buchmuller:1987ur}
  W.~Buchm\"uller, B.~Lampe and N.~Vlachos,
  Phys.\ Lett.\  B {\bf 197} (1987) 379.
%
\bibitem{Arzt:1994gp}
  C.~Arzt, M.~B.~Einhorn and J.~Wudka,
  Nucl.\ Phys.\  B {\bf 433} (1995) 41
  [hep-ph/9405214].
%
\bibitem{Grzadkowski:1997cj}
  B.~Grzadkowski, Z.~Hioki and M.~Szafra\'nski,
  Phys.\ Rev.\  D {\bf 58} (1998) 035002 
  [hep-ph/9712357].
%
\bibitem{Grzadkowski:1995te}
  B.~Grzadkowski,
  Acta Phys.\ Polon.\  B {\bf 27} (1996) 921 
  [hep-ph/9511279].
%
\bibitem{Abbott:1980zj}
  L.~F.~Abbott and M.~B.~Wise, 
  Phys.\ Rev.\  D {\bf 22} (1980) 2208.
%
\bibitem{Wilczek:1979hc}
  F.~Wilczek and A.~Zee,
  Phys.\ Rev.\ Lett.\  {\bf 43} (1979) 1571.
%
\bibitem{Leung:1984ni}
  C.~N.~Leung, S.~T.~Love and S.~Rao,
  Z.\ Phys.\  C {\bf 31} (1986) 433.
%
\bibitem{Iskrzynski:2010xxx}
M.~Iskrzy\'nski, M.$\,$Sc.$\;$Thesis, 
  {\it ``Classification of higher-dimensional operators in the Standard Model''},
University of Warsaw, 2010 (in Polish).
%
\bibitem{Iskrzynski:2010yyy}
  M.~Iskrzy\'nski, talk presented at the 17th IMPRS Workshop, Munich, July 2010,\\
  {\tt http://indico.mppmu.mpg.de/indico/conferenceDisplay.py?confId=901}~.
%
\bibitem{Grzadkowski:2008mf}
  B.~Grzadkowski and M.~Misiak,
  Phys.\ Rev.\  D {\bf 78} (2008) 077501
  [arXiv:0802.1413].
%
\bibitem{Grzadkowski:1995hi}
  B.~Grzadkowski and J.~Wudka,
  Phys.\ Lett.\  B {\bf 364} (1995) 49 
  [hep-ph/9502415].
%
\bibitem{AguilarSaavedra:2010zi}
  J.~A.~Aguilar-Saavedra,
  arXiv:1008.3562.
%
\bibitem{Hioki:2008XXX} Z.~Hioki, private communication, 2008.
%
\bibitem{Alonso:2014zka}
  R.~Alonso, H.~M.~Chang, E.~E.~Jenkins, A.~V.~Manohar and B.~Shotwell,
  Phys.\ Lett.\ B {\bf 734} (2014) 302
  [arXiv:1405.0486]. 
%
\end{thebibliography}
\end{document}